\documentclass{PoS}

\newcommand{\bea}{\begin{eqnarray}}
\newcommand{\eea}{\end{eqnarray}}

\title{Extending complex Langevin simulations to full QCD at nonzero density}

\ShortTitle{Extending complex Langevin simulations to full QCD at nonzero density}

\author{\speaker{D\'enes Sexty}  
\thanks{ I thank Gert Aarts, 
Erhard Seiler and Ion-Olimpiu Stamatescu for many discussions 
and collaboration on related work. 
A large part of the numerical calculations for this project was done on the 
bwGRiD (http://www.bw-grid.de), member of the German D-Grid initiative, 
funded by BMBF and MWFK Baden-W\"urttemberg.
}\\
        ITP, University of Heidelberg\\
        E-mail: \email{sexty.denes@gmail.com}}

      \abstract{\noindent Simulations of full QCD at nonzero 
   baryon density using light quark masses 
        are presented. The sign problem is evaded by the usage of the
        complex Langevin equation.  The simulations are 
stabilized by the gauge cooling procedure for small lattice spacings. 
 The method allows simulations at high densities, up to the saturation.
 The sign average is measured in the full as well as the phasequenched
        theory.  Results are compared to the HQCD approach, in which
        the spatial hopping terms of fermionic variables are dropped, and
        good agreement is found at large masses. 
                  }

\FullConference{31st International Symposium on Lattice Field Theory LATTICE 2013\\
		 July 29 - August 3, 2013\\
		 Mainz, Germany}

\begin{document}

\section{Introduction}

The determination of the phase 
diagram of finite density QCD is one of the great problems of 
theoretical physics today.
Lattice QCD is defined in terms of the euclidean path integral 
\bea \label{qcdZ}
Z = \int DU e^{-S_g[U]} \det M(\mu,U) 
\eea
where $S_g[U]$ is the Yang-Mills action of the gauge fields and
$M(\mu,U)$ is the Dirac-matrix of the quark fields. 
The well known sign problem prevents importance sampling 
simulations at $\mu \neq 0 $, as the determinant of the 
fermion matrix is a complex number in this case.
Various methods have been invented 
to circumvent the problem, but these are of limited use \cite{pdf}, mostly 
being applicable for $ \mu/T \lesssim 1 $. 
An exception is the complex Langevin method, which 
is not limited to small chemical potential.
 In this work I demonstrate that 
the algorithm can be successfully extended to full QCD with light 
quark masses on smooth lattices.

Shortly after the method of stochastic quantisation 
was proposed \cite{Parisi:1980ys},
it was noticed that it can be extended to 
complex actions \cite{parisi}. 
The Langevin equation gets complexified and the variables
are extended to a complexified manifold, which in the case
of $\textrm{SU}(N)$ variables is $\textrm{SL}(N,\mathbb{C})$.
The original theory is recovered for the 
averages of the analytically continued observables.
This approach can be applied in many areas, for example for 
the real time path integral 
problem \cite{bergesstam,bergessexty,opt}, or Yang-Mills theory 
with a $\Theta$-term.

A recent approach also relying on the complexification 
of the variable space is the Lefschetz-thimble approach  \cite{lefschetz},
 where the path of the integration is shifted off the real axis in 
the complex plane to a path where the phase of the 
measure is constant. A residual sign problem is still present
coming from the Jacobian of the transformation. 
This approach seems to be related to stochastic 
quantisation \cite{aartslefschetz}.

The complex Langevin method is known to sometimes produce convergent 
but wrong results, in other cases it solves theories 
with a severe sign problem.
It has been tested for various systems inspired by finite-density 
QCD \cite{aartsstam,Aarts:2008wh,Aarts:2010gr,aj,gaugecooling}, 
in some cases unsuccessfully \cite{Ambjorn:1985iw,Ambjorn:1986fz,aarts,Pawlowski:2013gag,Mollgaard:2013qra}.

Although a complete understanding of the successes and breakdowns is still
missing, our analytic understanding is improving steadily
 \cite{ajss,guralnik,haarmeas,Duncan:2013wm,Aarts:2013uza}, and 
one can gain an insight whether 
the results are trustworthy using requirements such as the fast decay of the 
distributions.

Recently an important addition to the simulation algorithm of gauge 
theories called 'gauge cooling' was proposed \cite{gaugecooling,Aarts:2013uxa}.
This procedure ensures that the distribution of the variables 
on the complex plane stays well localized, and this 
leads to correct results.
It was shown that with this addition the simulation 
of an approximation to QCD called HQCD (in which spatial hoppings 
of fermions are dropped) can be successfully simulated \cite{gaugecooling},
as well as full QCD with light quark masses \cite{fullqcd}.

\section{The complex Langevin equation for lattice QCD}

The discretised complex Langevin equation (CLE) for the 
link variables $U_{x,\nu}$ of lattice QCD 
with Langevin time step $\epsilon$ is written as 
\bea U_{x,\nu} (\tau+\epsilon) = 
  R_{x,\nu} (\tau) U_{x,\nu}(\tau),
\eea
with 
\bea
R_{x,\nu}(\tau) = \textrm{exp} \left[ i \sum\limits_a 
 \lambda_a ( \epsilon K_{ax\nu}  + \sqrt\epsilon \eta_{ax\nu} ) 
\right]
\eea
Here $\lambda_a$ are the generators of the gauge group, i.e. 
the Gell-Mann matrices. The drift force is determined as
\bea
K_{ax\nu}= -D_{ax\nu} S_\textrm{eff}[U]
\eea
from the effective action $S_\textrm{eff}[U] = S_g[U] + \textrm{ln det} M[\mu,U]$ with 
the left derivative
\bea
D_a f(U) = \left. \partial_\alpha f( e^{i \alpha \lambda_a} U ) \right|_{\alpha=0}.
\eea
In the case of QCD the drift term is written as 
\bea 
 K_{ax\nu} &=& -D_{ax\nu} S_g [U] 
 + {N_F\over 4 }\textrm Tr[ M^{-1}(\mu,U) 
  D_{ax\nu} M (\mu,U)  ].
\eea
For the calculation of the fermionic drift the bilinear 
noise scheme is used \cite{batrouni,fukugita1987}. The numerical cost is 
therefore one CG solution for every update. The number of CG iterations 
needed for a given precision is highly dependent 
on the chemical potential, with larger $\mu$ requiring more iterations.
 
This Langevin update is interspersed with several gauge cooling updates,
which make the simulation stable for not too small $\beta$ parameter
values \cite{gaugecooling,fullqcd}. Typically the main cost of the simulation
is the CG solution, the gauge force as well as the gauge cooling require 
less effort.


In the reweighting approach one defines the phasequenched theory
\bea \label{pqtheory}
Z_\textrm{pq} = \int DU e^{-S_g[U]} | \det M(\mu,U) | 
= \int DU e^{-S_g[U]} \det M(\mu_I,U) ,
\eea
where the complex phase of the determinant is dropped. This change can 
also be understood physically if we introduce an isospin 
dependent chemical potential $\mu_I$. We assume 
an even number of flavors. In this 
case isospin partners
will have the opposite phase of the determinant, and the total
determinant is positive, as written in eq. (\ref{pqtheory}), making
 importance sampling possible. This theory can also be simulated 
with the Langevin equation, where the change to the full
theory amounts to taking the real part 
of the drift force in the Langevin update. Gauge cooling is 
of course not needed in this case, one can reunitarize the link variables 
by hand to ensure they stay in the $\textrm{SU}(3)$ group 
in spite of the numerical rounding errors.

In this work I use the Wilson plaquette action 
for the gauge fields with $N_F=2$ or $N_F=4$ flavors of unimproved staggered 
fermions.

\section{Results}

We are interested in the standard observables such as 
the Polyakov loop and inverse Polyakov loop
\bea \label{poldef}
 P = {1\over V} \sum_x \textrm{Tr} P_x , \qquad 
 P' = {1\over V} \sum_x \textrm{Tr} P^{-1}_x \quad
\textrm{    with } \quad  P_x = \prod\limits_{\tau=0}^{N_T-1} U_{(\tau,x),4}, 
\eea
the chiral condensate
\bea \label{ccdef}
\langle \Psi^\dagger (x) \Psi(x) \rangle = \langle \partial \ln Z /\partial m \rangle / \Omega,
\eea
with $\Omega = V/ T$ the space-time volume, and the density of the fermions
\bea \label{ndef}
\langle n \rangle& =& {1\over \Omega} {\partial \ln Z \over \partial \mu },
\eea
measured in units of the saturation density. 
The average phase factor of the determinant is defined as 
\bea
\langle e^{2 i \varphi} \rangle = \left\langle 
{\det M( \mu) \over \det M(-\mu)} \right\rangle. 
\eea
This gives an impression of the distribution of the 
phases of the determinant. 
The average phase factor measured in the phasequenched theory is a direct 
indicator of the severeness of the sign-problem.

\begin{figure}
  \includegraphics[width= 0.5\columnwidth]{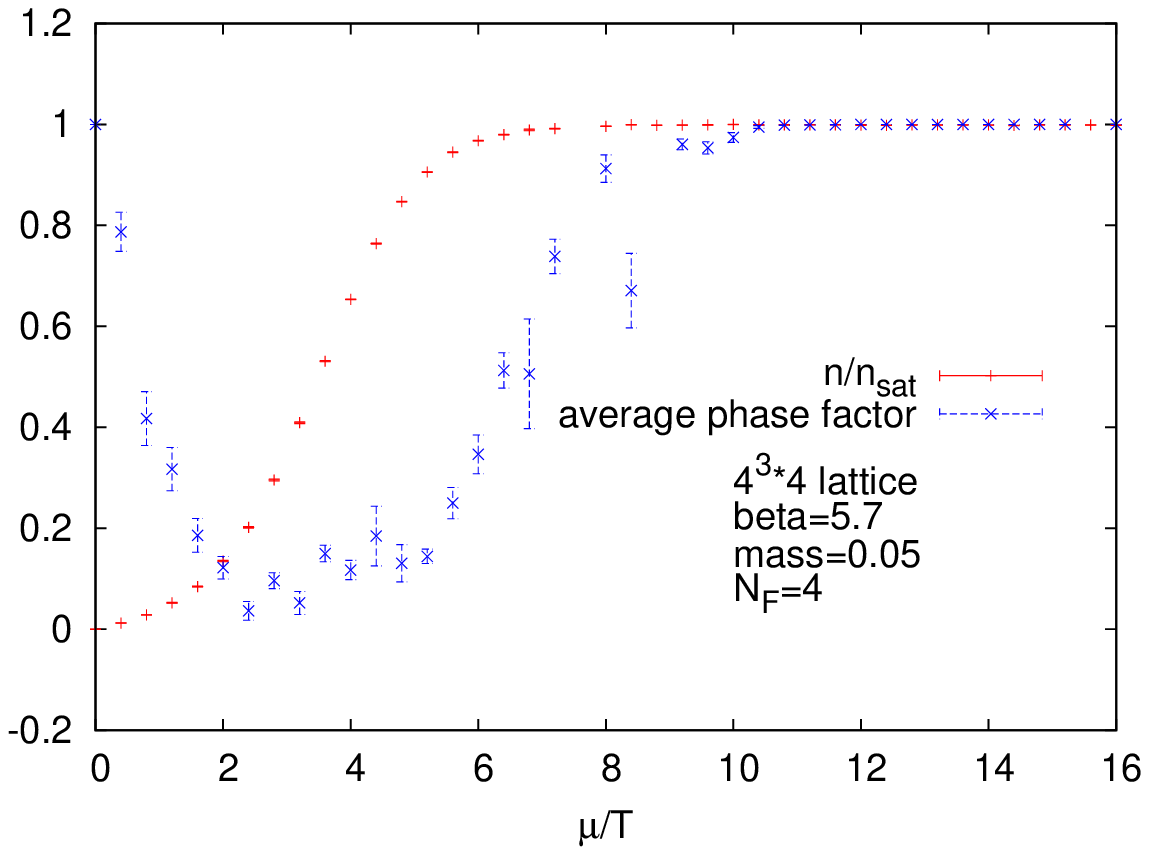}  
  \includegraphics[width= 0.5\columnwidth]{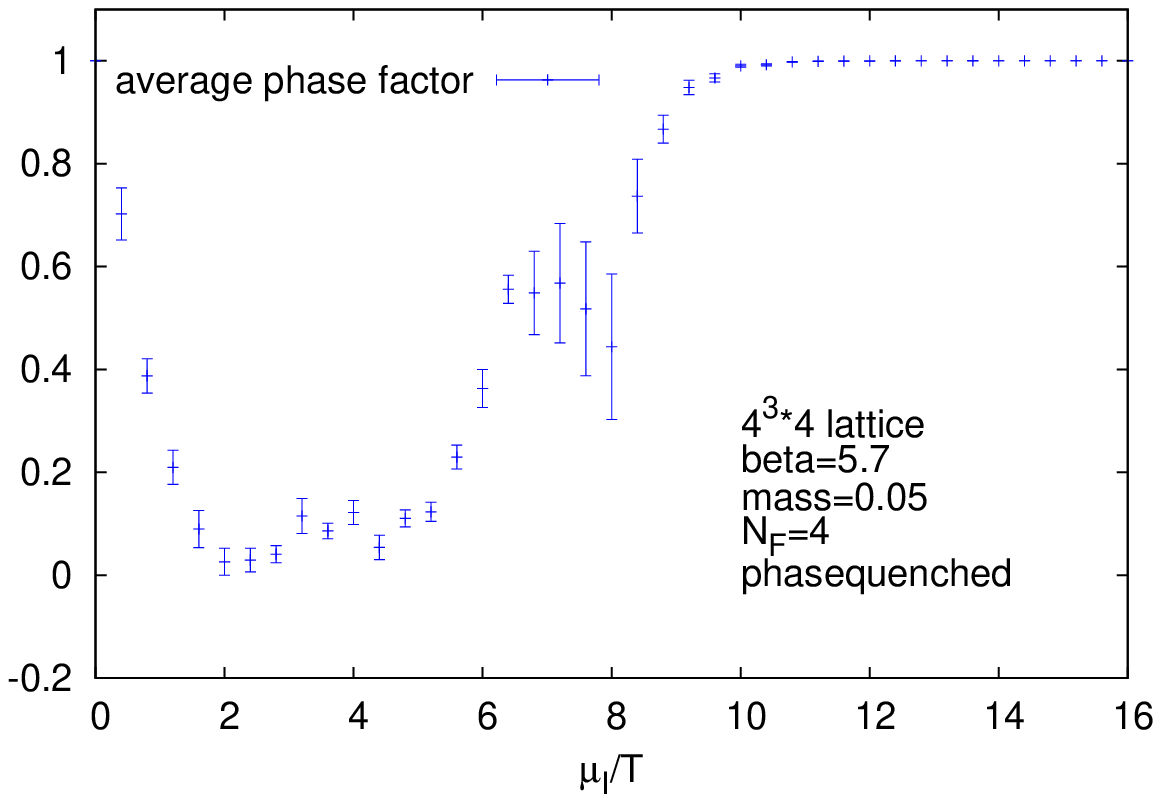}  
  \caption{ On the left panel the average phase factor and density 
is shown as a function 
of the chemical potential. On the right panel 
the average phase factor in the phasequenched theory is shown.}
\label{signdens}
\end{figure}

In Fig.~\ref{signdens} the density of the fermions as well as the 
average phase factor is shown. A small lattice is used 
as the direct calculation 
of the determinant is very costly. The parameters are such that  
the high temperature phase is tested, well above the phase 
transition (which is around $\beta_c\approx 5.04$ for the parameters used).
Therefore the density starts to increase right away, as there are available
excitations in the plasma. This is in contrast to the 
low temperature Silver Blaze behavior, where the chemical 
potential must reach a threshold before 
anything happens \cite{Cohen:2003kd,owe}. At high chemical potentials the 
saturation is reached, where all available fermionic 
states are filled. Note that the average phase factor is close to 
zero in a large region of chemical potentials in the full 
as well as the phasequenched theory. This makes the reweighting 
approach in this range of chemical potentials unfeasible.

\begin{figure}
\begin{center}
  \includegraphics[width= 0.55\columnwidth]{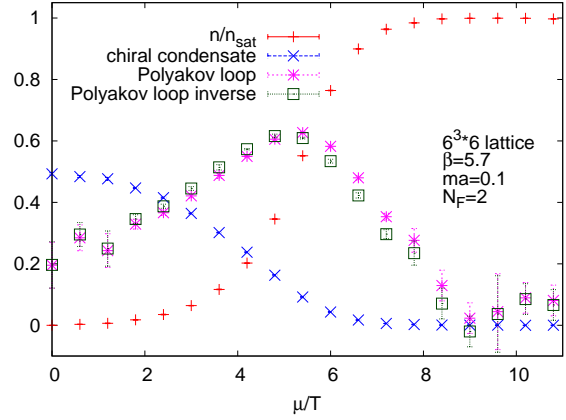}  
\end{center}
\caption{ The fermion density , 
the chiral condensate and the
trace of the Polyakov loop and its inverse as a function of 
the chemical potential. }
\label{horiz_slice}
\end{figure}

Fermionic observables as well as the Polyakov loops 
are measured at a fixed temperature as a function 
of the chemical potential in Fig.~\ref{horiz_slice}.
One observes the expected behavior of the Polyakov loops: as the fermionic
density increases, they grow as the system gets more ordered. The 
inverse Polyakov loop peaks before the 
Polyakov loop \cite{gaugecooling}.
Eventually, in the saturation region the Polyakov loops have to vanish,
 as the fermions have no influence on the system, 
and the $Z_3$ symmetry is restored. This is also reflected in the bigger 
fluctuations of the Polyakov loops.

\subsection{Comparison with HQCD}

\begin{figure}[!ht]
\begin{center} 
 \includegraphics[width= 0.55\columnwidth]{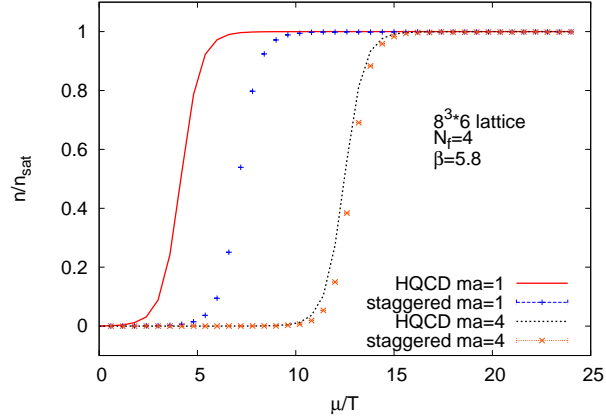}  
\end{center}
  \caption{ Comparison of the average densities 
 measured in HQCD and in full QCD with staggered fermions.}
\label{compplot1}
\end{figure}

\begin{figure}[!ht]
  \includegraphics[width= 0.5\columnwidth]{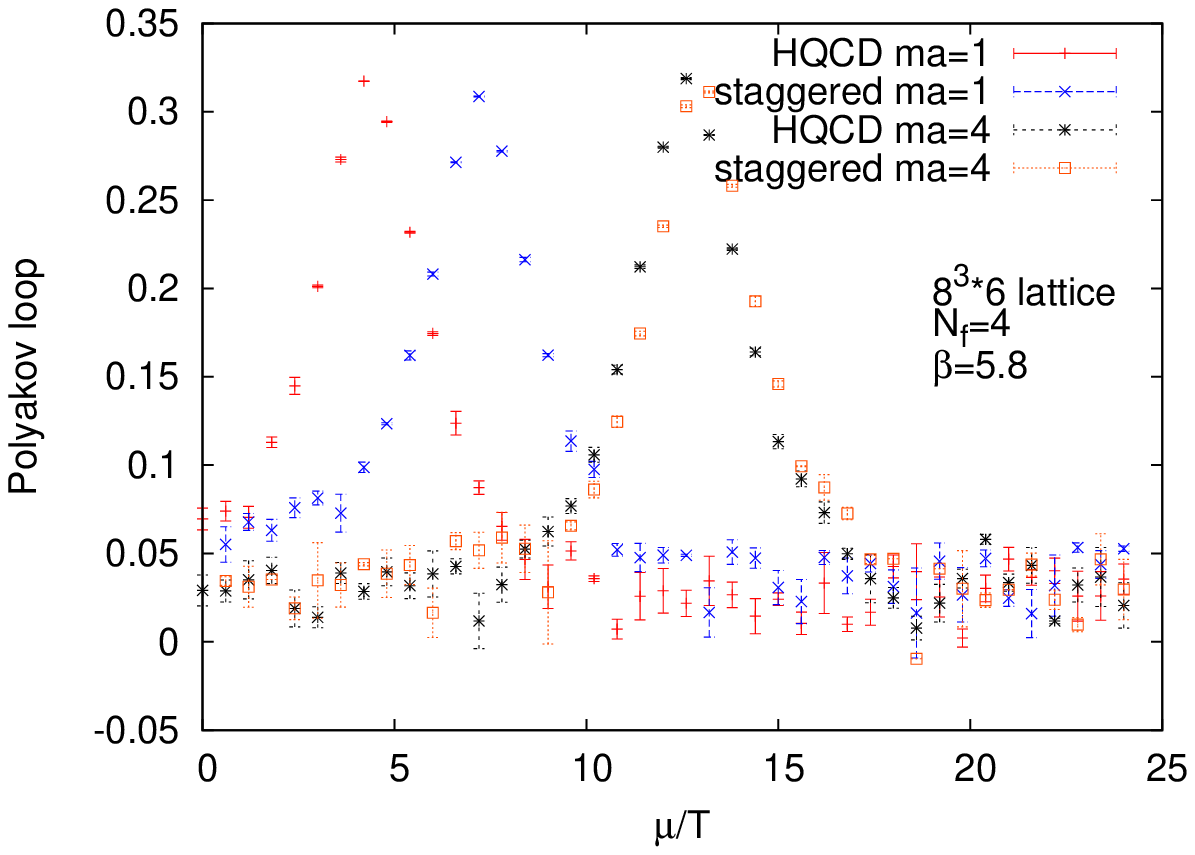}  
  \includegraphics[width= 0.5\columnwidth]{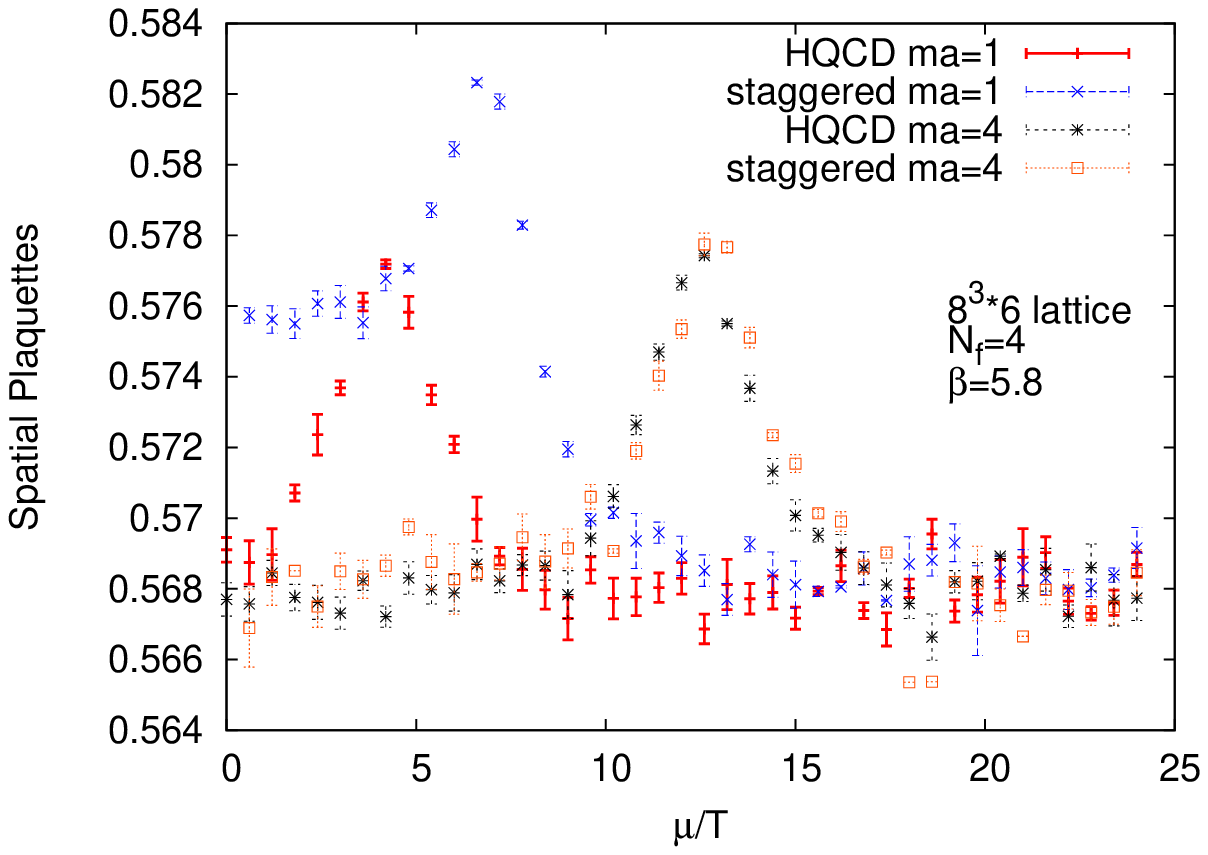}  
  \caption{ Comparison of the average Polyakov loops (left)
and average spatial plaquettes (right) 
 measured in HQCD and in full QCD with staggered fermions.}
\label{compplot2}
\end{figure}

A well known approximation to full QCD at high quark 
masses and large chemical potentials is the 
Heavy Quark QCD (HQCD) \cite{oldhqcd,feo}. In this approach 
the spatial hoppings are neglected, as they give 
a subleading contribution. The determinant 
simplifies to 
\bea
\det (M(\mu,U)) = \prod\limits_x \det (1+C P_x ) \det (1 + C' P^{-1}_x)
\eea
with the Polyakov loop (\ref{poldef})
and the parameters $ C= e^{\mu N_T} / (2 m)^{N_T} $ and 
$ C' = e^{-\mu N_T} / (2 m)^{N_T} $ with the staggered mass $m$. 
A similar approximation can also be carried out for Wilson fermions,
which was studied with the CLE in \cite{gaugecooling}. 

At large masses one can test the quality of the HQCD approximation
by comparing it to full QCD. In Fig.~\ref{compplot1} the 
fermion density is compared,
in Fig.~\ref{compplot2}, the Polyakov loops and the spatial plaquette 
variables are compared.


One notes that at $m=4$ the agreement becomes quantitatively good,
but the qualitative behavior is similar already at smaller mass,
with a 'rescaled' chemical potential. This does not 
prove that the results are fully reliable, but the HQCD approach has been
validated with reweighting at small $\mu$, therefore it increases
confidence that the CLE supplies correct results in this case. 

\section{Conclusions}

The complex Langevin equation has been applied to full QCD at 
non-zero density. The simulations are controlled using the 
gauge cooling procedure for not too big lattice spacings. 
The method allows calculations at arbitrary densities all the 
way up to saturation. There is no sign or overlap problem in complex 
Langevin simulations, the non-positiveness of the measure is 
taken into account by the analytic continuation of the variables
to the complexified manifold $\textrm{SL}(3,\mathbb{C})$.
The method allows simulations of full QCD with light fermions 
at high densities for the first time.

The procedure gives the physically expected results, the saturation
physics at large chemical potentials and the expected 
behavior of the Polyakov loops and inverse Polyakov loops is 
correctly reproduced.
To further increase confidence in the method, the next check would be 
to compare the full theory with the phasequenched one at low temperatures,
where the onset of the density has to reflect the crucial difference between 
the two theories.

\end{document}